\documentclass[aps,prb,twocolumn,showpacs,showkeys]{revtex4-1}
\usepackage{graphicx}%
\usepackage{dcolumn}
\usepackage{amsmath}   
\usepackage[usenames,dvipsnames]{pstricks}
\usepackage{epsfig}
\usepackage{pst-grad} 
\usepackage{pst-plot} 

\usepackage{graphicx} 
\usepackage{bm}
\usepackage{longtable}

\begin{document}

\title{Electrostatic fluctuations promote the dynamical transition in
  proteins} \author{Dmitry V.\ Matyushov}\email{dmitrym@asu.edu}
\affiliation{Center for Biological Physics, Arizona State University,
  PO Box 871604, Tempe, AZ 85287-1604 } \author{Alexander Y.\ Morozov}

\begin{abstract}
  Atomic displacements of hydrated proteins are dominated by phonon
  vibrations at low temperatures and by dissipative large-amplitude
  motions at high temperatures. A crossover between the two regimes is
  known as a dynamical transition. Recent experiments indicate a
  connection between the dynamical transition and the dielectric
  response of the hydrated protein. We analyze two mechanisms of the
  coupling between the protein atomic motions and the protein-water
  interface.  The first mechanism considers viscoelastic changes in
  the global shape of the protein plasticized by its coupling to the
  hydration shell. The second mechanism involves modulations of the
  motions of partial charges inside the protein by electrostatic
  fluctuations. The model is used to analyze mean square displacements
  of iron of metmyoglobin reported by M{\"o}ssbauer spectroscopy. We
  show that high flexibility of heme iron at physiological temperatures is
  dominated by electrostatic fluctuations. Two onsets, one arising
  from the viscoelastic response and the second from electrostatic
  fluctuations, are seen in the temperature dependence of the mean
  square displacements when the corresponding relaxation times enter
  the instrumental resolution window.
\end{abstract}

\maketitle

\section{Introduction}
\label{sec:1}
Measurements of the M{\"o}ssbauer absorption of $\mathrm{^{57}Fe}$ in
metmyoglobin crystals revealed that the mean square displacement
(msd) of this atom starts to grow faster with increasing temperature
above $T_D\simeq 200$ K.\cite{Parak:71} This finding was followed by
similar observations from neutron scattering,\cite{DosterNature:89}
which by now have been reported for a large number of proteins and
other biopolymers,\cite{Caliskan:06} all demonstrating the same
phenomenology.\cite{Gabel:02,Parak:03} The increase in the slope of
the protein msd as a function of temperature was called a ``dynamical
transition'', presently assigned to a rather broad range of onset
temperatures $T_D\simeq 200-240$ K. The basic observation is that the
high-temperature flexibility of the proteins much exceeds the linear
extrapolation of the low-temperature behavior characteristic of an
expanding solid. The low-temperature msd is well characterized by the
observed phonon spectrum of the protein,\cite{Parak:05} while the
high-temperature msd excess is linked to dissipative
long-wavelength modes with energies below $\simeq 3$
meV.\cite{Diehl:97,Achterhold:03,Marconi:08,Hinsen:08}

Early explanations of the dynamical transition offered scenarios
ranging from detrapping of the protein conformational motions from
low-energy states\cite{Frauenfelder:91,Zaccai:00,Parak:05} to the
glass transition in bulk water.\cite{DosterNature:89} Several recent
observations have shifted the focus to the protein hydration
shell. The disappearance of the dynamical transition in dry protein
powders,\cite{Kurkal:05} and its separate existence
for the hydration water,\cite{Wood:08,Chu:09}
point to a strong link between the dynamical transition and
the protein hydration shell.

Despite somewhat different semantics, the views in the field seem to
converge to the notion of the critical role of the hydration shell in
driving the dynamical transition. According to
Doster:\cite{DosterBBA:10} ``The onset of the dynamical transition
depends on the solvent viscosity near the protein surface.\dots The
protein-water $\alpha$-process consists of a concerted librational motion
of protein surface residues, coupled to translational jumps of water
molecules on the same time scale.'' The physical mechanism behind the
transition is assigned in this view to the caging of the protein by
water's hydrogen bonds, stiffening its conformational flexibility. As
the temperature increases, the population of broken hydrogen bonds
grows exponentially, resulting in a release of the protein
conformational flexibility in a narrow range of temperatures. Although
appealing, this concept does not address the key question of how
$ \sim 0.5-2$ ps events of hydrogen bond breaking develop into a $\sim 2\ \mu$s
collective relaxation process at $T_D$.

Frauenfelder and co-workers have recently suggested a somewhat
different scenario, also focusing on the dynamics of the protein
hydration shell.\cite{Fenimore:04,ChenPM:08,Frauenfelder:09} According
to their view, protein conformations are coupled to motions of the
hydration layer, with a relaxation time following an Arrhenius
law. This relaxation mode is therefore considered to be a secondary,
$\beta$-process according to the common classification adopted in the
field of supercooled liquids.\cite{Ediger:96} This secondary
relaxation is assessed by dielectric spectroscopy of protein samples
embedded into solid poly(vinyl)alcohol (PVA), thus eliminating the
bulk water relaxation from the dielectric
response.\cite{ChenPM:08,Frauenfelder:09}
 
The use of dielectric absorption of PVA-confined proteins yields a
surprisingly accurate account of the temperature dependence of the
Lamb-M{\"o}ssbauer factor equal to the fraction of recoilless
absorption, $f = \exp[ - k_0^2 \langle (\delta x)^2 \rangle ]$. Here,
$ \langle (\delta x)^2 \rangle$ is the msd of the heme iron of
metmyoglobin in projection on the wavevector $\mathbf{k}_0$ of
$\gamma$-radiation. The iron msd can be separated into a vibrational,
low-temperature component $ \langle (\delta q)^2 \rangle \propto T$,
described by the vibrational density of states (VDOS) of the protein,
and a high-temperature component $ \langle (\delta Q)^2 \rangle$
appearing at $T>T_D$.  Correspondingly, $f=f_q f_Q$ becomes a product
of two components, $f_q$ and $f_Q$.  It turns out that the temperature
variation of the Lamb-M{\"o}ssbauer factor $f_Q$ originating from the
high-temperature msd is exceptionally well described by the variance
of the sample dipole moment at the same hydration level
\begin{equation}
  \label{eq:1}
  f_Q =\exp[ -  k_0^2 \langle (\delta Q)^2 \rangle ]= \langle (\delta M)^2\rangle_{<}/  \langle (\delta M)^2\rangle .
\end{equation}
In this equation, $\langle (\delta M)^2\rangle_{<}$ is defined as the integral of the
frequency-dependent variance of the sample dipole moment $M_{\omega}$ over
the frequencies below (subscript ``$<$'') the instrumental frequency
$\omega_{\text{obs}}=1/ \tau_{\text{obs}}$, $\tau_{\text{obs}}=140$
ns.\cite{Frauenfelder:09} The parameter $f_Q$ then determines the
fraction of the sample dipole that has not had a chance to alter on
the life-time of the iron nucleus, thus keeping the nuclei in
resonance for M{\"o}ssbauer absorption. 

The empirical connection between the dynamical transition and dipolar
fluctuations is additionally supported by recent observations of
breaks in the dependence of the terahertz dielectric absorption on
temperature at typical values of $T_D$.\cite{HePRL:08} All these
observations, although advancing the field toward identifying the
physical modes responsible for high-temperature flexibility of
proteins, pose a significant conceptual challenge.

Both M{\"o}ssbauer and neutron-scattering techniques probe translational
atomic motions on their corresponding resolution windows. It seems
therefore natural to relate the break in the temperature dependence of
the msd to changes in the dynamic and/or static properties of atomic
translations. This is the conceptual framework behind the
glass-transition scenario\cite{DosterBBA:10} which, even in the
current form emphasizing the hydration layer, is focused on the caging
arrest, i.e., on the primary effect of short-ranged repulsive
interactions in the system.  On the contrary, the dielectric
measurements\cite{Frauenfelder:09,Khodadadi:08,HePRL:08} shift the
focus to the long-ranged electrostatic interactions, which is what
dielectric spectroscopy is sensitive to at the first place. What is
the correct view?

Answering this question requires gaining deeper insights into the
actual physical modes coupled to the protein msd at high temperatures,
the problem that has evaded direct experimental inquiry so
far. Numerical simulations point to enhanced fluctuations of hydrogen
bonds of hydration water at high
temperatures,\cite{Tarek_PhysRevLett:02} but those can be projected on
either density or dipolar collective modes. Since a strong link
between the dynamical transition and the hydration shell has been
clearly established, the main question posed by recent studies is
whether density or orientational collective modes drive the
transition.\cite{Singh:81} They are mostly decoupled by symmetry and
can therefore be considered as two distinct mechanisms of altering the
protein flexibility.

The goal of this paper is to present some initial estimates of the
relative importance of the density and orientational fluctuations in
the protein dynamical transition.  We model the density fluctuations
at the protein-water interface by viscoelastic response and the
orientational dipolar fluctuations in terms of electrostatic response.
The dependence of the onset temperature $T_D$ on the observation
window is an important ingredient of the
observations,\cite{DosterBBA:10,Magazu:10} which is introduced into
the theory by limiting the range of frequencies over which the
response functions are integrated,\cite{DMjcp1:09} similarly to Eq.\
(\ref{eq:1}). We start with formulating the model, followed by the
results of calculations.

\section{Model}
\label{sec:2}
The purpose of our model is to determine the msd of a single atom,
heme iron of metmyoglobin, as a function of temperature. The internal
motions of the protein can roughly be separated into two modes, the
phonon vibrations $\mathbf{q}$ and dissipative large-scale motions $\mathbf{Q}$. Even though
phonons do not formally exist in proteins, we will use this language
to distinguish short-wavelength vibrations from motions altering the
protein's global shape. Accordingly, we will split the coordinate of the iron atom
$\mathbf{r}=\mathbf{q}+\mathbf{Q}$ into two statistically independent
components, $\mathbf{q}$ and $\mathbf{Q}$. The former can be expanded
in protein's normal modes or directly calculated from the VDOS
measured, for instance, by the phonon-assisted M{\"o}ssbauer
scattering.\cite{Achterhold:02,Parak:05,Leu:08} The corresponding msd
of the vibrational coordinate $\mathbf{q}$ is, in the classical limit,
a linear function of temperature%
\begin{equation}
  \label{eq:2}
  \langle (\delta q)^2\rangle = a_q T .
\end{equation}
The proportionality coefficient $a_q$ is calculated from the VDOS
according to the standard prescriptions.

The dissipative motions of the protein are seen in neutron scattering
spectra as a quasielastic peak with energies below $\simeq 4$ meV,
growing in intensity with increasing temperature.\cite{Marconi:08}
With the sound velocity in a protein\cite{Achterhold:03,Leu:10} of
$\simeq 1700$ m/s, the wavelength of the corresponding vibrations is
about 26 \AA, which is comparable with the diameter of myoglobin, $2R
= 36$ \AA. These modes therefore alter the global shape of
the protein, which is the domain of the viscoelastic response.

In order to obtain a first-order estimate of the geometry changes
involved, we will model the protein motions as radial viscoelastic
vibrations of a sphere of radius $R$ immersed in a viscoelastic water
continuum. The low-frequency variance of the iron coordinate is then
simply related to the radius fluctuations of the sphere as $\langle
(\delta Q^2)\rangle = (r/R)^2 \langle (\delta R)^2\rangle $. The
latter can be found by solving the standard equations of
viscoelasticity\cite{Landau7,Christensen:03} yielding the response
function $\chi_R(\omega)$ connecting the change of the sphere's radius
$R$ to an oscillatory pressure $p(t)=p_0e^{i\omega t}$ applied to the
sphere's surface. The result is\cite{Landau7}
\begin{equation}
  \label{eq:3}
  \chi_R(\omega) = - \frac{1}{4\pi R} \frac{1}{3\Delta K_p(\omega) + 4 \mu_w(\omega)} .
\end{equation}
Here, $\Delta K_p(\omega) = K_p(\omega) - K_{p,0}$ is the viscoelastic bulk modulus of
the protein minus its bulk modulus $K_{p,0}$ at zero
frequency. Further, $\mu_w(\omega)$ is the shear modulus of water. Applying
the fluctuation-dissipation theorem\cite{Boon:91} to Eq.\ (\ref{eq:3}), one
gets
\begin{equation}
  \label{eq:4}
  \langle (\delta Q_{\omega})^2 \rangle = - \frac{2k_{\text{B}} T r^2}{3\omega V_p}\mathrm{Im} \frac{1}{3\Delta K_p(\omega)
    + 4 \mu_w(\omega)} ,  
\end{equation}
where $V_p$ is the protein volume.

We now proceed to calculating the Lamb-M{\"o}ssbauer
factor\cite{Singwi:60,Cranshaw:85} 
\begin{equation}
  \label{eq:5}
  f = \langle \left| \langle e^{ik_0 x} \rangle \right|^2 \rangle_{\text{het}},
\end{equation}
where $x = \mathbf{\hat k}_0 \cdot (\mathbf{q} + \mathbf{Q})$ is the
projection of the iron displacement on the direction of photon
propagation, $\mathbf{\hat k}_0=\mathbf{k}_0/k_0$. There are two
averages in this definition: the inner angular brackets denote an
ensemble average over the protein and water modes affecting the
position of iron in a single protein, while the outer angular brackets
denote the average over the proteins in the sample.  This second
average carries the subscript ``het'' to emphasize that it reflects
the heterogeneity of the sample, such as for instance variations in
the hydration level among different proteins in the protein powder. We
do not consider the heterogeneous average in our present study and
limit ourselves by the inner average only. This approximation amounts,
in experimental techniques, to considering the narrow feature of the
M{\"o}ssbauer absorption line and subtracting the broad base-line
originating from the sample heterogeneity.\cite{Parak:05} Accordingly,
the experimental points shown in Fig.\ \ref{fig:1} are obtained from
the area  $f(T)$ of the narrow line as $-\ln f(T)/(k_0)^2$, $k_0=53.2$
\AA$^{-1}$.

The ensemble average over the water/protein statistical
distribution can be described in terms of the free energy $F(x)$ such
that the inner brackets in Eq.\ (\ref{eq:5}) become
\begin{equation}
  \label{eq:6}
   \langle e^{ik_0 x} \rangle  = Z^{-1} \int e^{ik_0 x - \beta F(x)} dx,
\end{equation}
where $\beta = 1/(k_{\text{B}}T)$ and $Z=\int \exp[-\beta F(x)] dx$.
The free energy $F(x)$ is determined by
projecting\cite{ChaikinLubensky} the manifold of $\mathbf{q}$ and
$\mathbf{Q}$ coordinates on the single coordinate $x$
\begin{equation}
  \label{eq:7}
\begin{split}
  e^{-\beta F(x)}  =  \int & \delta [ x - \mathbf{\hat k}_0 \cdot (\mathbf{q} + \mathbf{Q})]\\
                   & \langle e^{-\beta H_0(\mathbf{q},\mathbf{Q}) -\beta z
                 \phi_w(\mathbf{q},\mathbf{Q})} \rangle_w\ 
               d\mathbf{q}d\mathbf{Q} . 
\end{split}
\end{equation}
In this equation, $H_0(\mathbf{q},\mathbf{Q})$ is the Hamiltonian of
classical harmonic vibrations of the protein and $\phi_w$ is the
electrostatic potential of the surrounding dielectric medium acting on
the heme iron carrying charge $z$. The subscript ``w''  in Eq.\ \eqref{eq:7}
specifies water as the main source of the electrostatic
fluctuations. It is not, however, required by the theory, and slow
protein motions, not included in the calculation of $\langle(\delta
q)^2\rangle$, can contribute to the fluctuations of the electrostatic
potential as well (see below).  

The Hamiltonian $H_0(\mathbf{q},\mathbf{Q})$ can be given in the
Gaussian form
\begin{equation}
  \label{eq:8}
  \beta H_0(\mathbf{q},\mathbf{Q}) = \frac{\delta q^2}{2\langle (\delta q)^2\rangle } + 
                                        \frac{\delta Q^2}{2\langle (\delta Q)^2\rangle_{>}
                                        }, 
\end{equation}
where the variance of $\mathbf{q}$ is given by Eq.\ (\ref{eq:2}) and
the variance of $\mathbf{Q}$ requires additional explanation.  

The
limited instrumental time $\tau_{\text{obs}}$ affects the observables
and, in fact, the dynamical transition itself becomes possible only
when the relaxation time of a collective mode of the hydration shell
coupled to high-temperature protein's motions enters the experimental
observation window.\cite{Frauenfelder:09,DosterBBA:10} Therefore, the
variance of the slow dispersive motions of the heme iron is not a
thermodynamic variable referring to an infinite observation window,
but a property affected by instrumental
resolution.\cite{Knapp:83,DMjcp1:09} This is reflected by the
subscript ``$>$'' in Eq.\ (\ref{eq:8}) which indicates that $\langle
(\delta Q)^2\rangle_{>}$ is calculated by integrating the response
function in Eq.\ (\ref{eq:4}) over the frequencies exceeding the
observation frequency $\omega_{\text{obs}}=\tau_{\text{obs}}^{-1}$
\begin{equation}
  \label{eq:9}
  \langle (\delta Q)^2\rangle_{>} = \int_{\omega_{\text{obs}}}^{\infty} \langle (\delta Q_{\omega})^2\rangle (d\omega/ \pi) . 
\end{equation}

The statistical average over the electrostatic fluctuations can be
simplified by a first-order expansion of the potential $\phi_w$ in
$x$: $\phi_w(\mathbf{q},\mathbf{Q}) \simeq \phi_{w,0} - x E_w$, where
$\phi_{w,0}$ is the potential at the equilibrium position of iron and
$E_w$ is the electric field projected on $\mathbf{\hat k}_0$.
Assuming that $E_w$ is a Gaussian variable, one gets a Gaussian form
of $\beta F(x)= x^2/(2\langle(\delta x)^2 \rangle )$ where the
variance $\langle(\delta x)^2 \rangle$ is
\begin{equation}
  \label{eq:11}
  \langle(\delta x)^2 \rangle =  \langle(\delta x)^2 \rangle_{\text{el}}/M_E .
\end{equation}
Here, the elastic msd $ \langle(\delta x)^2 \rangle_{\text{el}}=
\langle (\delta q)^2\rangle +\langle (\delta Q)^2\rangle_{>}$ is the
sum of two statistically decoupled components. Further, the correction
$M_E$ represents softening of atomic vibrations by electrostatic
fluctuations of the hydration shell. 

Similarly to the viscoelastic effect, the electrostatic softening
depends on the observation window.  Accounting again for the frequency
cutoff introduced by the finite instrumental resolution, it is given
in the form
\begin{equation}
  \label{eq:12}
  M_E = 1 - (\beta z)^2 \langle(\delta x)^2 \rangle_{\text{el}}
             \int_{\omega_{\text{obs}}}^{\infty} C_E(\omega)d\omega/(2\pi) ,
\end{equation}
where $C_E(\omega)$ is the Fourier transform of the time autocorrelation
function of the field $E_w(t)$
\begin{equation}
  \label{eq:13}
  C_E(\omega) = \int_{-\infty}^{\infty} \langle \delta E_w(t) \delta E_w(0)\rangle e^{i\omega t} dt .
\end{equation}

By applying the fluctuation-dissipation theorem\cite{Boon:91} once
again, one can recast Eq.\ (\ref{eq:12}) in terms of the response
function $\chi_E(\omega)$ representing the polar response to an
oscillating dipole probe $m(t)= m_0\exp(i\omega t)$ placed at the
position of the iron atom
\begin{equation}
  \label{eq:14}
   M_E = 1 - \beta z^2 \langle(\delta x)^2 \rangle_{\text{el}}
             \int_{\omega_{\text{obs}}}^{\infty} \chi_E''(\omega)d\omega/(\pi \omega ) .
\end{equation}
This form of the correction factor accounting for the dipolar
fluctuations of the water shell is used in the calculations below.

\section{Mean square displacement}
\label{sec:3}
Here we outline the calculations performed using the model developed
in this paper. We need to mention that many parameters entering the
model are not experimentally available. Some of them can be
potentially extracted from numerical simulations. The usefulness of
simulations is, however, limited for interpreting the experimental
data since reproducing heterogeneous conditions of partially hydrated protein
powders presents significant challenges to simulation protocols.
Likewise, the viscoelastic model used here should be viewed as only a
first step toward a more realistic description of the elastic response
of hydrated proteins. However, one of the major conclusions of this
paper is the dominance of electrostatics in the high-temperature
flexibility of proteins and a relatively small effect of viscoelastic
motions on the iron msd. This observation puts high priority to the
development of the electrostatic component of the model, and makes the
limitations of the modeling of the viscoelastic response less
critical.

\begin{figure}
  \centering
  \includegraphics*[width=7cm]{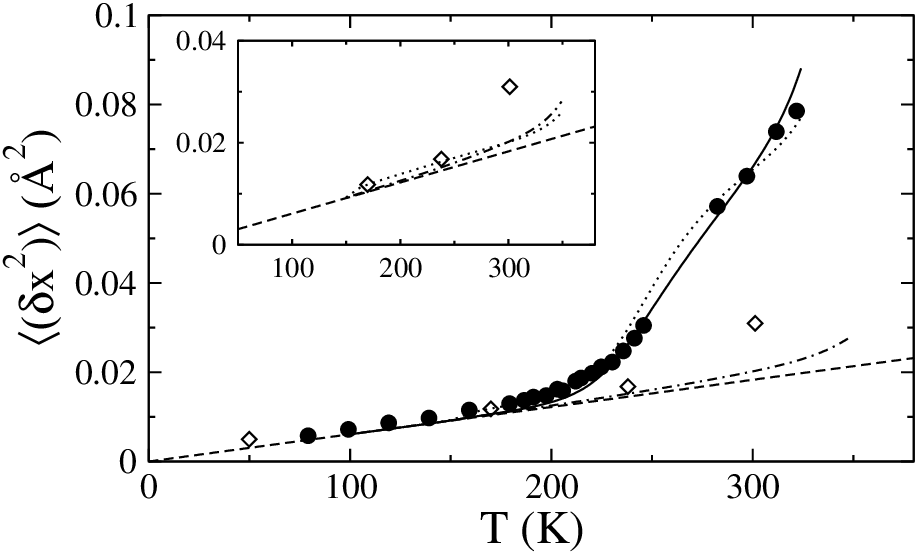}
  \caption{Temperature dependence of the msd of iron in
    metmyoglobin. Experimental msd\cite{Parak:05} are shown by
    circles. Diamonds refer to the msd calculated from the VDOS
    measured from phonon-assisted M{\"o}ssbauer effect at different
    temperatures,\cite{Parak:05} whereas the dashed line is the same
    calculation from the VDOS at 235 K.\cite{Achterhold:02} The
    dash-dotted line refers to the msd combining phonons with
    viscoelastic oscillations of the protein shape and the solid line
    refers to the combination of all three effects: protein's phonons,
    viscoelastic shape oscillations, and dipolar fluctuations of the
    hydration layer. The solid line is obtained with $\Delta K_p(T)$ as
    explained in the text, while the dotted lines refer to the
    temperature-independent $\Delta K_p=3.2$ GPa obtained from the fit of
    Eq.\ (\ref{eq:18}) to the protein intrinsic compressibility at 298
    K.\cite{Mori:06} The inset shows the rise of the msd due to the
    elastic motions near the glass transition temperature of the
    protein, $T_g\simeq 180 \pm 15$ K.}
  \label{fig:1}
\end{figure}

The viscoelastic response functions entering Eq.\ (\ref{eq:3}) were
taken in the Maxwell form\cite{Boon:91}
\begin{equation}
  \label{eq:15}
\begin{split}
  \Delta K_p(\omega) &= \frac{\Delta K_p i \omega \tau_p(T)}{1+ i\omega \tau_p(T)}, \\
   \mu_w(\omega)& = \frac{G_{\infty}i\omega\tau_w(T)}{1+i\omega\tau_w(T)} .
\end{split}
\end{equation}
In this equation, $\Delta K_p=K_{p,\infty}-K_{p,0}$ is the change in the bulk
protein modulus between infinite and zero frequencies and $G_{\infty}$ is
the high-frequency shear modulus of water; $\tau_{p,w}(T)$ are the
corresponding relaxation times. From the Maxwell equation, one gets
the shear viscosity $\eta_w(T)= G_{\infty}(T)\tau_w(T)$ which is well tabulated
down to the water nucleation temperature.\cite{Harris:04} The shear
relaxation time was obtained from $\eta_w(T)$ and $G_{\infty}(T)/
\mathrm{GPa}= 1.68 - 0.0127(T-273)$ taken from Ref.\
\onlinecite{Slie:66}. The resulting $\tau_w(T)$ turns out to be close
to the exponential relaxation time of the longitudinal modulus
extracted from inelastic x-ray scattering:\cite{Monaco:99} $\tau_{\ell}(T)/
\mathrm{s} =0.84 \times 10^{-15} \exp(1910\ \mathrm{K}/T)$.

The VDOS of metmyoglobin powders is well established by
phonon-assisted M{\"o}ssbauer
measurements,\cite{Achterhold:02,Achterhold:03,Parak:05} and the
temperature slope $a_q$ in Eq.\ (\ref{eq:2}) was calculated from the
experimental VDOS (dashed line in in Fig.\ \ref{fig:1}).  Figure
\ref{fig:1} also shows a rise in the msd due to the onset of
viscoelastic oscillations of the protein (dash-dotted line). These
results were obtained by adopting Eq.\ (\ref{eq:15}) for the elastic
moduli with the protein's relaxation time $\tau_p(T)$ from the
measurements done on dry protein powders.\cite{Khodadadi:08} The
relaxation process in dry proteins is too slow to enter the
observation window of the spectrometer and $\Delta K_p(\omega) \simeq
\Delta K_p$ in Eq.\ (\ref{eq:15}).

This notion implies that that the frequency dependence of the moduli,
e.g., Debye vs dispersive relaxation does not significantly affect the
outcome of the calculations and only $\Delta K_p(T)$ matters for the
temperature dependence of the viscoelastic msd.  The latter was
adopted to reproduce the experimental intrinsic
compressibility\cite{Kharakoz:00} of myoglobin\cite{Mori:06} $\beta_T =
11.04$ Mbar$^{-1}$ at $T=298$ K and the temperature variation of
Young's moduli of myoglobin crystals at lower
temperatures\cite{Morozov:85,Morozov:93} (see below). The use of
experimental compressibility to parametrize the model also implies
that the overall viscoelastic component of atomic displacements is
limited by the thermodynamic experimental value and can only be lower
for a given instrumental observation window.

The most uncertain part of our analysis is the calculation of the
electric field response function $\chi_E(\omega)$. It can be
calculated by solving the Poisson equation for the heme immersed in
the heterogeneous dielectric formed by the protein and its hydration
layer. However, the assignment of the dielectric constants to both the
protein and the thin layer of water surrounding it in the powder is
subject to significant uncertainties.  We will therefore restrict
ourselves to rough estimates aimed to establish whether electrostatic
fluctuations can produce a significant effect on the msd under a
reasonable set of assumptions. The results of fitting are additionally
supported by Molecular Dynamics simulations of the fully hydrated
metmyoglobin.\cite{DMunpubl:11}

The dielectric response to charges immersed in a polar medium is
dominated by longitudinal modes of dipolar
polarization\cite{DMjcp1:04} characterized by the dielectric modulus
$\epsilon(\omega)^{-1}$, where $\epsilon(\omega)$ is the complex,
frequency-dependent dielectric constant of the protein-water
mixture. The dielectric response function $\chi_E(\omega)$ establishes
the reaction field of the dielectric medium in response to a probe
dipole placed at the position of the heme iron. It scales as the
inverse cube of the characteristic size $d$ of the heme. The loss
function $\chi_E''(\omega)$ can therefore be written in the
form
\begin{equation}
  \label{eq:10}
  \chi_E''(\omega) = \frac{1}{d^3}\ \frac{\epsilon''(\omega)}{|\epsilon(\omega)|^2} .
\end{equation}

\begin{figure}
  \centering
  \includegraphics*[width=7cm]{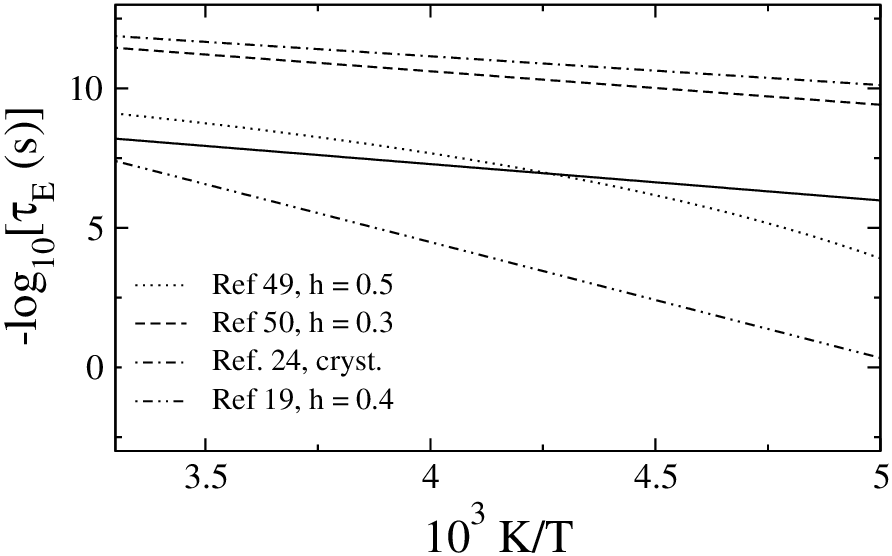}
  \caption{Dielectric longitudinal relaxation time $\tau_L(T)
    =(\epsilon_{\infty}/ \epsilon_s) \tau_E(T)$ reported in the
    literature\cite{Bonura:10,Schiro:09,Frauenfelder:09} with the
    protein hydration levels indicated in the legend. The results from
    Ref.\ \cite{Frauenfelder:09} refer to myoglobin confined in
    PVA, results from Ref.\ \cite{Singh:81} are for metmyoglobin
    crystals, and Refs.\ \cite{Bonura:10,Schiro:09} refer to
    metmyoglobin powders. The solid line refers to the longitudinal,
    Debye relaxation time of the reaction field response function
    $\chi_E(\omega)$ [Eq.\ (\ref{eq:16})], $\tau_L(T)/ \mathrm{s}=3.2 \times
    10^{-13}\exp[3000\ \mathrm{K}/T]$, used to fit the experimental
    msd for the heme iron.\cite{Parak:05} 
}
  \label{fig:2}
\end{figure}

Dielectric properties of partially hydrated proteins have not been
well characterized since the results are strongly affected by both the
sample preparation and the hydration level. Even dielectric relaxation
times measured on samples of close hydration level are rather
inconsistent. This point is illustrated in Fig.\ \ref{fig:2} where
results on partially hydrated myoglobin powders of hydration level
$h=0.3-0.5$ (in g of water per g of protein) have been
assembled.\cite{Swenson:06,Frauenfelder:09,Schiro:09,Bonura:10} We
also show measurements done on myoglobin crystals\cite{Singh:81} for
the sake of comparison. Multiple relaxation processes are common for such
measurements, and the fastest relaxation, commonly attributed to the
hydration shell,\cite{Khodadadi:08,Schiro:09} is shown in Fig.\
\ref{fig:2}.

Even if we could firmly establish the proper relaxation time for the
the sample dipole moment, this would not necessarily give us the
relaxation time of the reaction field correlation function required
for $\chi''_E(\omega)$. Because of the linear scaling of the dipole moment variance
with the number of dipoles, dielectric measurements emphasize the
effect of outer solvation shells, while $\chi''_E(\omega)$ is dominated by
waters closest to the probe dipole (heme's iron). In view of these
uncertainties, we have constructed the temperature-dependent
relaxation time $\tau_E(T)$ that, together with the parameter $d$ in Eq.\
(\ref{eq:10}), allows us to fit the experimental msd. 

Assuming the Debye form for the fast relaxation component in
$\epsilon(\omega)$,\cite{Khodadadi:08,Bonura:10} the response function
in Eqs.\ (\ref{eq:14}) and (\ref{eq:10}) gains the form
\begin{equation}
  \label{eq:16}
  \beta z^2\chi_E''(\omega) = \frac{1}{\delta^2} \ \frac{\omega \tau_L}{1+(\omega\tau_L)^2},
\end{equation}
where the parameter $\delta$, $\delta^{-2} = \beta z^2 c_0/ d^3$ sets
up a charatecteristic length and $c_0=\epsilon_{\infty}^{-1} -
\epsilon_s^{-1}$ is the Pekar factor. In addition,
$\tau_L=(\epsilon_{\infty}/ \epsilon_s) \tau_E$ is the longitudinal
dielectric time and $\epsilon_{\infty}$ and $\epsilon_s$ are the
high-frequency and static dielectric constants, respectively. The
parameters $\delta=0.12$ \AA\ and $\tau_L(T)/ \mathrm{s}=3.2 \times
10^{-13} \exp[3000\ \mathrm{K}/T]$ (solid line in Fig.\ \ref{fig:2})
were used in fitting the experimental msd.  The fitting relaxation
time $\tau_L(T)$ is generally consistent with dielectric measurements
and, in addition, the Arrhenius slope of $\tau_L(T)$ matches our
simulations of the protein Stokes-shift dynamics at elevated
temperatures.\cite{DMjpcb2:08} More detailed calculations might
require replacing one-relaxation Debye dynamics in Eq.\ (\ref{eq:16})
with dispersive dynamics characterized by a distribution of relaxation
time, as suggested by the NMR experiment.\cite{Lusceac:10}

\begin{figure}
  \centering
  \includegraphics*[width=7cm]{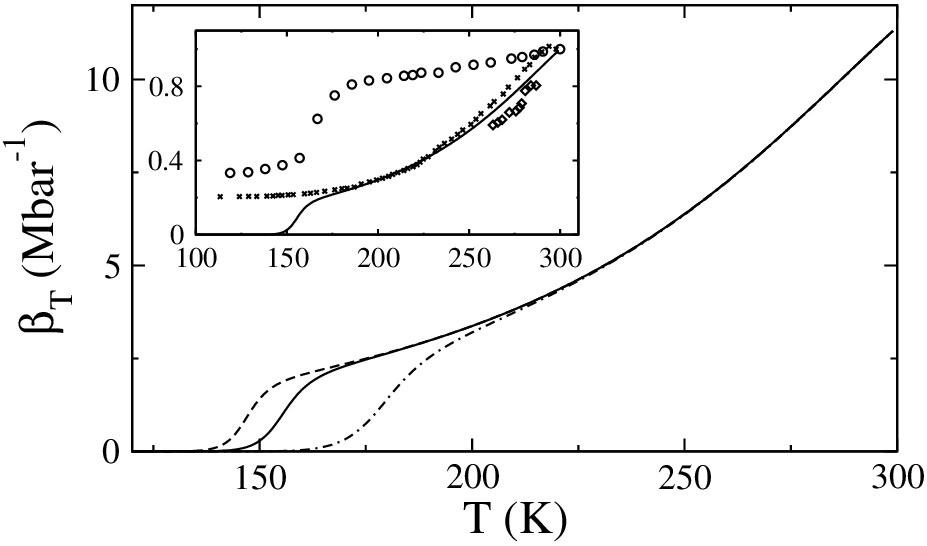}
  \caption{Intrinsic isothermal compressibility $\beta_T(T)$ of a protein
    calculated from Eq.\ (\ref{eq:18}) with $\omega_{\text{obs}}=1$ (dashed
    line), $10^{2}$ (solid line), and $10^{3}$ MHz (dash-dotted line).
    The protein and water parameters are those adopted for the
    calculation of the myoglobin msd, with $\Delta K_p(T) / \mathrm{GPa}=
    3.22 - 0.03\times (T - 298) + 0.00025\times (T - 298)^2$ obtained to fit the
    protein intrinsic compressibility at 298 K\cite{Mori:06} and the
    temperature dependence of Young's moduli (crosses in the inset).
    The inset shows experimental expansivity of the hydration shell of
    lysozyme (circles),\cite{Doster:10} experimental inverse Young's
    moduli of myoglobin crystals exposed to air of 95--100\%
    (diamonds)\cite{Morozov:93} and 75\% humidity
    (crosses).\cite{Morozov:85} The solid line shows $\beta_T(T)$
    calculated from Eq.\ (\ref{eq:18}) at $\omega_{\text{obs}}=10^{2}$ MHz;
    all curves are normalized to the corresponding values at $T=298$
    K. }
  \label{fig:3}
\end{figure}

The overall fluctuations of the protein volume are determined by the
response function in Eq.\ \eqref{eq:3} combining the dynamic elastic
moduli of the protein and the hydration shell. This connection can be
used to parameterize the model on volumetric properties of hydrated
proteins, in particular on protein's intrinsic
compressibility.\cite{Kharakoz:00} For a given instrumental
resolution, one obtains from Eq.\ (\ref{eq:3}) for the isothermal
compressibility $\beta_T\propto \langle(\delta V_p)^2\rangle$ of the
protein
\begin{equation}
  \label{eq:18}
  \beta_T = -(6/ \pi) \int_{\omega_{\text{obs}}}^{\infty} \mathrm{Im}\left[ 3\Delta
    K_p(\omega) + 4\mu_w(\omega) \right]^{-1} (d\omega / \omega) .
\end{equation}

In Fig.\ \ref{fig:3} we show $\beta_T(T)$ for the parameters adopted in
the calculations of the iron msd and several values of $\omega_{\text{obs}}$.
The intrinsic compressibility of the protein rises sharply at the
point close to protein's glass transition $T_g$. The latter depends on the
observation window, but is close to reported values $T_g\simeq 180\pm 15$
K\cite{Morozov:93,Doster:10,Khodadadi:10} marked by breaks in several
observable parameters.\cite{Khodadadi:10} The rise of compressibility
at $T_g$ in our calculations is caused by the water component of the
viscoelastic response function when the relaxation time $\tau_w$ becomes
smaller than $\tau_{\text{obs}}$. This result is consistent with the
glass transition of the hydration shell expansivity\cite{Doster:10}
shown in the inset in Fig.\ \ref{fig:3}.  The inset in Fig.\
\ref{fig:3} also shows compressibilities obtained from experimentally
reported Young's moduli of myoglobin
crystals\cite{Morozov:85,Morozov:93} assuming that their Poisson
ratios are independent of temperature. The temperature variation of $\Delta
K_p(T)$ in our calculations shown in Fig.\ \ref{fig:1} was chosen to
match these data.

The fit of $\Delta K_p(T)$ to crystalline Young's moduli results in an
upward increase of the elastic msd at the highest temperatures shown
in Fig.\ \ref{fig:1}. This upward increase reflects pre-melting of
myoglobin crystals when their Young's moduli approach
zero.\cite{Morozov:93} Since the melting temperature is typically
higher in protein powders,\cite{Morozov:93} $\Delta K_p(T)$ obtained
from fitting the crystal data might overestimate these effects; the
iron msd with a temperature-independence $\Delta K_p=3.2$ GPa is shown
by the dotted line in Fig.\ \ref{fig:1}.

\begin{figure}
  \centering
  \includegraphics*[width=8cm]{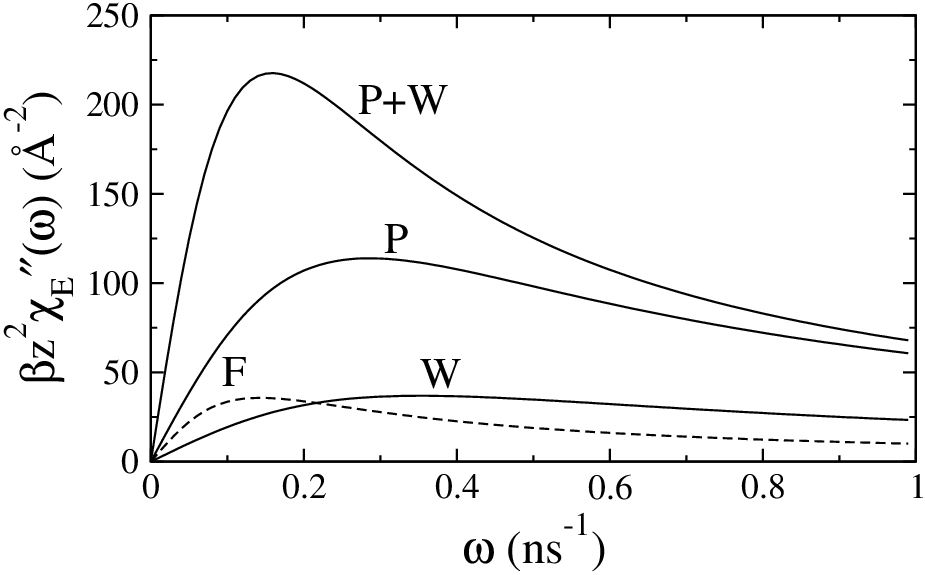}
  \caption{The loss function $\beta z^2 \chi_E''(\omega)$ obtained from the
    fitting of the experimental msd to Eqs.\ (\ref{eq:11}),
    (\ref{eq:14}), and (\ref{eq:16}) (marked as ``F'') and from direct
    MD simulations of the electric field acting on the iron of
    metmyoglobin.  The plot shows the result for the overall electric
    field produced by protein and water (``P+W'') and by protein
    (``P'') and water (``W'') separately. The simulation trajectories
    ($T=300$ K) were 45 ns long, 35 ns of which  were used for
    collecting the correlation functions.\cite{DMunpubl:11} }
  \label{fig:4}
\end{figure}

The results shown in Fig.\ \ref{fig:1} indicate that electrostatic
fluctuations far outweigh viscoelastic vibrations in the iron msd. We
additionally confirm this outcome by comparing the function $\beta z^2
\chi_E''(\omega)$ from our fitting to the same function obtained from
MD simulation of the fully hydrated metmyoglobin.\cite{DMunpubl:11}
Figure \ref{fig:4} shows the response functions from the electric
field fluctuations produced by the protein and water combined and by
each component separately. The height of the maximum quantifies the
strength of the msd modulation by the corresponding electrostatic
component, and it is of main importance for this comparison.

We have assumed so far that the protein is electrostatically
non-polar, and its hydration shell is the main source of the
electrostatic fluctuations. It does not need to be so. Low-frequency
motions, not included in the VDOS used to caculate $\langle(\delta
q)^2\rangle$, can modulate the protein's partial charges (dipole
moments of $\alpha$-helices, ionized surface residues, etc.)  and
compete in the elecrostatic noise with the hydration layer. This might
be particularly true for partially hydrated protein powders where the
fluctuations of the water dipoles are probably reduced to motions of
polarized domains around ionized surface residues.

Figure \ref{fig:4} in fact shows that the protein component of
$\chi_E''(\omega)$ exceeds that of water, and its maximum is higher
than that of the fitting function from Eq.\ \eqref{eq:16}.  The
electrostatic fluctuations of the protein itself are therefore
sufficient to produce the observable msd and, in addition, our
estimates do not seem to overestimate the effect of the electrostatic
fluctuations on the msd. The primary role of water in powders might be
reduced to ionizing the surface residues of the protein and
plasticizing its motions above $T_g$ (Fig.\ \ref{fig:3}).  Water in
patches solvating ionized residues is strongly coupled to the protein
both electrostatically and by surface hydrogen bonds. The relaxation
times of their electrostatic response functions are therefore close
(Fig.\ \ref{fig:4}), resulting in matching onset temperatures of the
dynamical transition for each component.\cite{Wood:08,Chu:09}

Further, the overall loss function $\chi_E''(\omega)$, which includes
cross-correlations between the water and protein electric fields,
shows a slower relaxation time than each component separately. The
relaxation time of 6.3 ns of the essentially Debye overall function
$\chi_E''(\omega)$ is close to $\tau_L(300\ \mathrm{K})\simeq 7$ ns
adopted in fitting of the experimental msd. It is this loss function,
combining the protein and water electrostatics, that is of primary
interest for the modeling of the high-temperature flexibility of
proteins.

\section{Discussion}
\label{sec:4}
The picture presented here assigns an increase in the protein msd at
the dynamical transition to the entrance of a collective relaxation
time of the protein-water interface into the observation window of the
spectrometer.\cite{Daniel:02,Khodadadi:08,Frauenfelder:09} We consider
two types of interfacial fluctuations, elastic modes changing the
global shape of the protein and electrostatic
fluctuations. Electrostatics turn out to be the main factor affecting
the high-temperature portion of the msd. 

The longitudinal relaxation time of the electric field fluctuations,
$\tau_L(T)$, determines the transition temperature by the condition
$\omega_{\text{obs}}\tau_L(T_D)\simeq 1$. With the Arrhenius form for
the relaxation time $\tau_L(T)$, this condition predicts a logarithmic
dependence of $T_D$ on the observation frequency,
\begin{equation}
  \label{eq:17}
  T_D \propto \left|\ln[\omega_{\text{obs}} \tau_0 ]\right|^{-1} ,
\end{equation}
where $\tau_0$ is the preexponent in $\tau_L(T)$.  For instance, with
the observation window of neutron scattering of $\simeq 500$ ps and of
M{\"o}ssbauer spectroscopy of 140 ns, the above equation yields 1.4
for the ratio of $T_D$ values measured by neutron and M{\"o}ssbauer
techniques ($\tau_0=10^{-13}$ s). This estimate assumes equal
electrostatic relaxation times for (mostly surface) protons and heme
iron, which is likely not true. The actual picture is also more
complex as several slope changes contribute to the overall temperature
dependence of the msd.\cite{Krishnan:08} It is also the case with the
present model producing two different onsets arising from viscoelastic
and electrostatic fluctuations. An increase in $T_D$ was also reported
for proteins solvated in glycerol and in concentrated sucrose-water
solutions.\cite{DosterBBA:10} Although an increase in viscosity does
shift $T_D$ in the right direction according to Eq.\ (\ref{eq:17}),
the alteration of the effective polarity of the hydration layer and
the surface charge distribution of the protein might be other factors
contributing to the shift. Generally, the the present model predicts a
decrease in the protein atomic displacements for hydration in solvents of
lower polarity.

\begin{figure}
  \centering
  \includegraphics*[width=7cm]{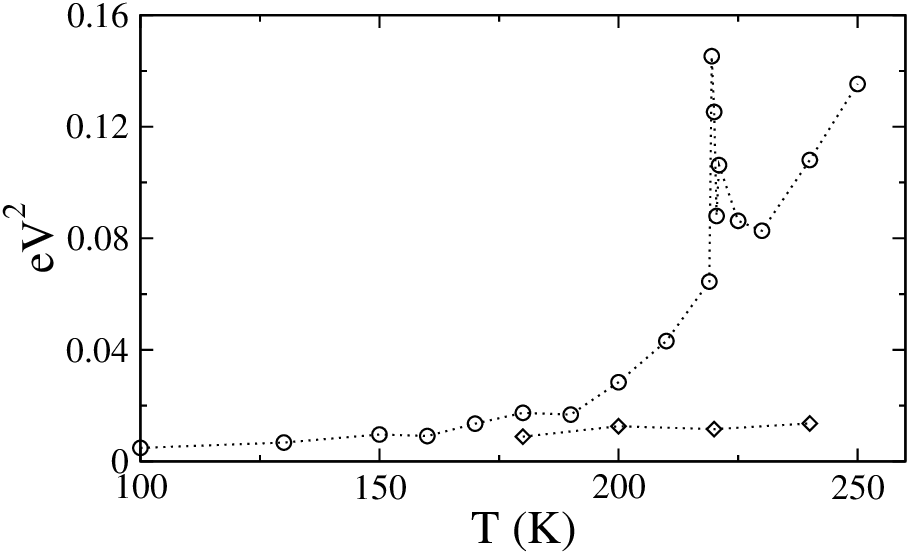}
  \caption{Variance of the water's electrostatic potential at the
    active site, $\Delta q^2 \langle (\delta \phi )^2\rangle $ of the protein plastocyanin from
    MD simulations (circles).\cite{DMpre2:08} Diamonds show the
    difference of water potentials in equilibrium with the active site
    carrying charges $q_1$ and $q_2$, $\beta^{-1} \Delta q(\langle \phi\rangle_1 - \langle \phi\rangle_2)$,
    $\Delta q=q_2 - q_1$.  This latter quantity is sensitive to
    high-frequency ballistic modes of the hydration water, but not to
    collective fluctuations of the shell dipole.\cite{DMjpcb:10} The
    two calculations coincide in the linear response approximation,
    which is valid at low temperatures. Linear response breaks down
    when the collective mode of water's dipolar polarization enters
    the observation window fixed by the length of the simulation
    trajectory.  The spike at $\simeq 220$ K in the potential variance
    carries signatures of a weak first-order transition, but its
    origin is currently unclear. }
  \label{fig:5}
\end{figure}

The main physical question looming behind the phenomenon of the
dynamical transition is what are the mechanisms and physical modes
allowing high flexibility of proteins at physiological
temperatures. We emphasize here electrostatic fluctuations as the
primary origin of the increase in the protein's atomic
displacements. This mechanism connects the translational manifold of
the protein's interior to the dipolar orientational manifold of the
hydration layer. While this connection was established empirically by
experiment [Eq.\ (\ref{eq:1})], numerical simulations directly show
the same basic phenomenology for the electrostatic fluctuations and
the atomic msd.

Figure \ref{fig:5} shows the results of numerical simulations for the
variance of the electrostatic potential produced by the water
hydration shell at the active site of the protein
plastocyanin.\cite{DMpre2:08} A break in the temperature dependence at
$T_D$ refers to the time-scale of $\simeq 10$ ns fixed by the length
of the simulation trajectory. The difference of the first moments of
the potential in the two redox states of the protein (diamonds in
Fig.\ \ref{fig:5}) gives the component of the same property produced
by the ballistic dynamics of the hydration shell and not sensitive to
its collective relaxation.\cite{DMjpcb:10} In a sense, the diamonds in
Fig.\ \ref{fig:5} are analogs to the diamonds in Fig.\ \ref{fig:1}
referring to the vibrational component of the msd.\cite{Parak:05}
There is a clear qualitative similarity between laboratory and
numerical results presented in Figs.\ \ref{fig:1} and \ref{fig:5}.

Because the response function of the water's electric field scales as
$d^{-3}$ with the distance $d$ from the surface inside the protein,
dipolar fluctuations of the hydration shell will mostly affect
protein's surface residues. The vibrations of the surface protons will
therefore be softer than of interior protons, and they will
stronger contribute to the observable msd. There is also a possibility
of ``surface melting'' when $M_E=0$ in Eq.\ (\ref{eq:11}) is reached
with rising temperature for a group of atoms. The low-temperature
conformation of the corresponding residues will become unstable, with
the instability released through a conformational transition.

The present model predicts higher flexibility for atoms carrying
higher partial charges. In case of heme iron this implies higher
flexibility of the protein oxidized state compared to the reduced
state. While this prediction qualitatively agrees with
experiment,\cite{Frolov:97,Jorgenson:05} more detailed studies are
required to distinguish the effect of electrostatic fluctuations from
the alteration of the VDOS also occurring upon changing the redox
state.

\section{Conclusions}
The model proposed here treats high-temperature atomic displacements
of the protein as a combination of viscoelastic deformation of the
global protein shape and electrostatic fluctuations coupled to the
atomic charge. We suggest that electrostatic fluctuations dominate the
high-temperature flexibility of proteins.

\acknowledgments This research was supported by the National Science
Foundation (CHE-0910905). We are grateful to Alexei Sokolov, Jan
Swenson, and Guo Chen for communicating results of their dielectric
measurements to us. DVM has greatly benefited from useful discussions
with Robert Young and Alexei Sokolov.

\bibliography{chem_abbr,dielectric,dm,statmech,protein,liquids,solvation,dynamics,glass,elastic}

\begin{thebibliography}{10}%
\makeatletter
\providecommand \@ifxundefined [1]{%
 \ifx #1\undefined \expandafter \@firstoftwo
 \else \expandafter \@secondoftwo
\fi
}%
\providecommand \@ifnum [1]{%
 \ifnum #1\expandafter \@firstoftwo
 \else \expandafter \@secondoftwo
\fi
}%
\providecommand \enquote [1]{``#1''}%
\providecommand \bibnamefont  [1]{#1}%
\providecommand \bibfnamefont [1]{#1}%
\providecommand \citenamefont [1]{#1}%
\providecommand\href[0]{\@sanitize\@href}%
\providecommand\@href[1]{\endgroup\@@startlink{#1}\endgroup\@@href}%
\providecommand\@@href[1]{#1\@@endlink}%
\providecommand \@sanitize [0]{\begingroup\catcode`\&12\catcode`\#12\relax}%
\@ifxundefined \pdfoutput {\@firstoftwo}{%
 \@ifnum{\z@=\pdfoutput}{\@firstoftwo}{\@secondoftwo}%
}{%
 \providecommand\@@startlink[1]{\leavevmode\special{html:<a href="#1">}}%
 \providecommand\@@endlink[0]{\special{html:</a>}}%
}{%
 \providecommand\@@startlink[1]{%
  \leavevmode
  \pdfstartlink
   attr{/Border[0 0 1 ]/H/I/C[0 1 1]}%
   user{/Subtype/Link/A<</Type/Action/S/URI/URI(#1)>>}%
  \relax
 }%
 \providecommand\@@endlink[0]{\pdfendlink}%
}%
\providecommand \url  [0]{\begingroup\@sanitize \@url }%
\providecommand \@url [1]{\endgroup\@href {#1}{\urlprefix}}%
\providecommand \urlprefix [0]{URL }%
\providecommand \Eprint[0]{\href }%
\@ifxundefined \urlstyle {%
  \providecommand \doi [1]{doi:\discretionary{}{}{}#1}%
}{%
  \providecommand \doi [0]{doi:\discretionary{}{}{}\begingroup
  \urlstyle{rm}\Url }%
}%
\providecommand \doibase [0]{http://dx.doi.org/}%
\providecommand \Doi[1]{\href{\doibase#1}}%
\providecommand \bibAnnote [3]{%
  \BibitemShut{#1}%
  \begin{quotation}\noindent
    \textsc{Key:}\ #2\\\textsc{Annotation:}\ #3%
  \end{quotation}%
}%
\providecommand \bibAnnoteFile [2]{%
  \IfFileExists{#2}{\bibAnnote {#1} {#2} {\input{#2}}}{}%
}%
\providecommand \typeout [0]{\immediate \write \m@ne }%
\providecommand \selectlanguage [0]{\@gobble}%
\providecommand \bibinfo [0]{\@secondoftwo}%
\providecommand \bibfield [0]{\@secondoftwo}%
\providecommand \translation [1]{[#1]}%
\providecommand \BibitemOpen[0]{}%
\providecommand \bibitemStop [0]{}%
\providecommand \bibitemNoStop [0]{.\EOS\space}%
\providecommand \EOS [0]{\spacefactor3000\relax}%
\providecommand \BibitemShut [1]{\csname bibitem#1\endcsname}%
\bibitem{Parak:71}%
  \BibitemOpen
  \bibfield{author}{%
  \bibinfo {author} {\bibfnamefont{F.}~\bibnamefont{Parak}}\ and\ \bibinfo
  {author} {\bibfnamefont{H.}~\bibnamefont{Formanek}},\ }%
  \bibfield{journal}{%
  \bibinfo {journal} {Acta Crystallogr. A}\ }%
  \textbf{\bibinfo {volume} {27}},\ \bibinfo {pages} {573} (\bibinfo {year}
  {1971})%
  \bibAnnoteFile{NoStop}{Parak:71}%
\bibitem{DosterNature:89}%
  \BibitemOpen
  \bibfield{author}{%
  \bibinfo {author} {\bibfnamefont{W.}~\bibnamefont{Doster}}, \bibinfo {author}
  {\bibfnamefont{S.}~\bibnamefont{Cusack}},\ and\ \bibinfo {author}
  {\bibfnamefont{W.}~\bibnamefont{Petry}},\ }%
  \bibfield{journal}{%
  \bibinfo {journal} {Nature}\ }%
  \textbf{\bibinfo {volume} {337}},\ \bibinfo {pages} {754} (\bibinfo {year}
  {1989})%
  \bibAnnoteFile{NoStop}{DosterNature:89}%
\bibitem{Caliskan:06}%
  \BibitemOpen
  \bibfield{author}{%
  \bibinfo {author} {\bibfnamefont{G.}~\bibnamefont{Caliskan}}, \bibinfo
  {author} {\bibfnamefont{R.~M.}\ \bibnamefont{Briber}}, \bibinfo {author}
  {\bibfnamefont{D.}~\bibnamefont{Thirumalai}}, \bibinfo {author}
  {\bibfnamefont{V.}~\bibnamefont{Garcia-Sakai}}, \bibinfo {author}
  {\bibfnamefont{S.~A.}\ \bibnamefont{Woodson}},\ and\ \bibinfo {author}
  {\bibfnamefont{A.~P.}\ \bibnamefont{Sokolov}},\ }%
  \bibfield{journal}{%
  \bibinfo {journal} {J. Am. Chem. Soc.}\ }%
  \textbf{\bibinfo {volume} {128}},\ \bibinfo {pages} {32} (\bibinfo {year}
  {2006})%
  \bibAnnoteFile{NoStop}{Caliskan:06}%
\bibitem{Gabel:02}%
  \BibitemOpen
  \bibfield{author}{%
  \bibinfo {author} {\bibfnamefont{F.}~\bibnamefont{Gabel}}, \bibinfo {author}
  {\bibfnamefont{D.}~\bibnamefont{Bicout}}, \bibinfo {author}
  {\bibfnamefont{U.}~\bibnamefont{Lehnert}}, \bibinfo {author}
  {\bibfnamefont{M.}~\bibnamefont{Tehei}}, \bibinfo {author}
  {\bibfnamefont{M.}~\bibnamefont{Weik}},\ and\ \bibinfo {author}
  {\bibfnamefont{G.}~\bibnamefont{Zaccai}},\ }%
  \bibfield{journal}{%
  \bibinfo {journal} {Quat.\ Rev.\ Biophys.}\ }%
  \textbf{\bibinfo {volume} {35}},\ \bibinfo {pages} {327} (\bibinfo {year}
  {2002})%
  \bibAnnoteFile{NoStop}{Gabel:02}%
\bibitem{Parak:03}%
  \BibitemOpen
  \bibfield{author}{%
  \bibinfo {author} {\bibfnamefont{F.~G.}\ \bibnamefont{Parak}},\ }%
  \bibfield{journal}{%
  \bibinfo {journal} {Rep.\ Prog.\ Phys.}\ }%
  \textbf{\bibinfo {volume} {66}},\ \bibinfo {pages} {103} (\bibinfo {year}
  {2003})%
  \bibAnnoteFile{NoStop}{Parak:03}%
\bibitem{Parak:05}%
  \BibitemOpen
  \bibfield{author}{%
  \bibinfo {author} {\bibfnamefont{F.~G.}\ \bibnamefont{Parak}}\ and\ \bibinfo
  {author} {\bibfnamefont{K.}~\bibnamefont{Achterhold}},\ }%
  \bibfield{journal}{%
  \bibinfo {journal} {J.\ Phys.\ Chem.\ Solids}\ }%
  \textbf{\bibinfo {volume} {66}},\ \bibinfo {pages} {2257} (\bibinfo {year}
  {2005})%
  \bibAnnoteFile{NoStop}{Parak:05}%
\bibitem{Diehl:97}%
  \BibitemOpen
  \bibfield{author}{%
  \bibinfo {author} {\bibfnamefont{M.}~\bibnamefont{Diehl}}, \bibinfo {author}
  {\bibfnamefont{W.}~\bibnamefont{Doster}}, \bibinfo {author}
  {\bibfnamefont{W.}~\bibnamefont{Petry}},\ and\ \bibinfo {author}
  {\bibfnamefont{H.}~\bibnamefont{Schober}},\ }%
  \bibfield{journal}{%
  \bibinfo {journal} {Biophys. J.}\ }%
  \textbf{\bibinfo {volume} {73}},\ \bibinfo {pages} {2726} (\bibinfo {year}
  {1997})%
  \bibAnnoteFile{NoStop}{Diehl:97}%
\bibitem{Achterhold:03}%
  \BibitemOpen
  \bibfield{author}{%
  \bibinfo {author} {\bibfnamefont{K.}~\bibnamefont{Achterhold}}\ and\ \bibinfo
  {author} {\bibfnamefont{F.~G.}\ \bibnamefont{Parak}},\ }%
  \bibfield{journal}{%
  \bibinfo {journal} {J. Phys.: Condens. Matter}\ }%
  \textbf{\bibinfo {volume} {15}},\ \bibinfo {pages} {S1683} (\bibinfo {year}
  {2003})%
  \bibAnnoteFile{NoStop}{Achterhold:03}%
\bibitem{Marconi:08}%
  \BibitemOpen
  \bibfield{author}{%
  \bibinfo {author} {\bibfnamefont{M.}~\bibnamefont{Marconi}}, \bibinfo
  {author} {\bibfnamefont{E.}~\bibnamefont{Cornicchi}}, \bibinfo {author}
  {\bibfnamefont{G.}~\bibnamefont{Onori}},\ and\ \bibinfo {author}
  {\bibfnamefont{A.}~\bibnamefont{Paciaroni}},\ }%
  \bibfield{journal}{%
  \bibinfo {journal} {Chem. Phys.}\ }%
  \textbf{\bibinfo {volume} {345}},\ \bibinfo {pages} {224} (\bibinfo {year}
  {2008})%
  \bibAnnoteFile{NoStop}{Marconi:08}%
\bibitem{Hinsen:08}%
  \BibitemOpen
  \bibfield{author}{%
  \bibinfo {author} {\bibfnamefont{K.}~\bibnamefont{Hinsen}}\ and\ \bibinfo
  {author} {\bibfnamefont{G.~R.}\ \bibnamefont{Kneller}},\ }%
  \bibfield{journal}{%
  \bibinfo {journal} {Proteins}\ }%
  \textbf{\bibinfo {volume} {70}},\ \bibinfo {pages} {1235} (\bibinfo {year}
  {2008})%
  \bibAnnoteFile{NoStop}{Hinsen:08}%
\bibitem{Frauenfelder:91}%
  \BibitemOpen
  \bibfield{author}{%
  \bibinfo {author} {\bibfnamefont{H.}~\bibnamefont{Frauenfelder}}, \bibinfo
  {author} {\bibfnamefont{S.~G.}\ \bibnamefont{Sligar}},\ and\ \bibinfo
  {author} {\bibfnamefont{P.~G.}\ \bibnamefont{Wolynes}},\ }%
  \bibfield{journal}{%
  \bibinfo {journal} {Science}\ }%
  \textbf{\bibinfo {volume} {254}},\ \bibinfo {pages} {1598} (\bibinfo {year}
  {1991})%
  \bibAnnoteFile{NoStop}{Frauenfelder:91}%
\bibitem{Zaccai:00}%
  \BibitemOpen
  \bibfield{author}{%
  \bibinfo {author} {\bibfnamefont{G.}~\bibnamefont{Zaccai}},\ }%
  \bibfield{journal}{%
  \bibinfo {journal} {Science}\ }%
  \textbf{\bibinfo {volume} {288}},\ \bibinfo {pages} {1604} (\bibinfo {year}
  {2000})%
  \bibAnnoteFile{NoStop}{Zaccai:00}%
\bibitem{Kurkal:05}%
  \BibitemOpen
  \bibfield{author}{%
  \bibinfo {author} {\bibfnamefont{V.}~\bibnamefont{Kurkal}}, \bibinfo {author}
  {\bibfnamefont{R.~M.}\ \bibnamefont{Daniel}}, \bibinfo {author}
  {\bibfnamefont{J.~L.}\ \bibnamefont{Finney}}, \bibinfo {author}
  {\bibfnamefont{M.}~\bibnamefont{Tehei}}, \bibinfo {author}
  {\bibfnamefont{R.~V.}\ \bibnamefont{Dunn}},\ and\ \bibinfo {author}
  {\bibfnamefont{J.~C.}\ \bibnamefont{Smith}},\ }%
  \bibfield{journal}{%
  \bibinfo {journal} {Chem. Phys.}\ }%
  \textbf{\bibinfo {volume} {317}},\ \bibinfo {pages} {267} (\bibinfo {year}
  {2005})%
  \bibAnnoteFile{NoStop}{Kurkal:05}%
\bibitem{Wood:08}%
  \BibitemOpen
  \bibfield{author}{%
  \bibinfo {author} {\bibfnamefont{K.}~\bibnamefont{Wood}}, \bibinfo {author}
  {\bibfnamefont{A.}~\bibnamefont{Fr{\"o}lich}}, \bibinfo {author}
  {\bibfnamefont{A.}~\bibnamefont{Paciaroni}}, \bibinfo {author}
  {\bibfnamefont{M.}~\bibnamefont{Moulin}}, \bibinfo {author}
  {\bibfnamefont{M.}~\bibnamefont{H{\"a}rtlein}}, \bibinfo {author}
  {\bibfnamefont{G.}~\bibnamefont{Zaccai}}, \bibinfo {author}
  {\bibfnamefont{D.~J.}\ \bibnamefont{Tobias}},\ and\ \bibinfo {author}
  {\bibfnamefont{M.}~\bibnamefont{Weik}},\ }%
  \bibfield{journal}{%
  \bibinfo {journal} {J. Am. Chem. Soc.}\ }%
  \textbf{\bibinfo {volume} {130}},\ \bibinfo {pages} {4586} (\bibinfo {year}
  {2008})%
  \bibAnnoteFile{NoStop}{Wood:08}%
\bibitem{Chu:09}%
  \BibitemOpen
  \bibfield{author}{%
  \bibinfo {author} {\bibfnamefont{X.-Q.}\ \bibnamefont{Chu}}, \bibinfo
  {author} {\bibfnamefont{A.}~\bibnamefont{Faraone}}, \bibinfo {author}
  {\bibfnamefont{C.}~\bibnamefont{Kim}}, \bibinfo {author}
  {\bibfnamefont{E.}~\bibnamefont{Fratini}}, \bibinfo {author}
  {\bibfnamefont{P.}~\bibnamefont{Baglioni}}, \bibinfo {author}
  {\bibfnamefont{J.~B.}\ \bibnamefont{Leao}},\ and\ \bibinfo {author}
  {\bibfnamefont{S.-H.}\ \bibnamefont{Chen}},\ }%
  \bibfield{journal}{%
  \bibinfo {journal} {J.\ Phys.\ Chem. B}\ }%
  \textbf{\bibinfo {volume} {113}},\ \bibinfo {pages} {5001} (\bibinfo {year}
  {2009})%
  \bibAnnoteFile{NoStop}{Chu:09}%
\bibitem{DosterBBA:10}%
  \BibitemOpen
  \bibfield{author}{%
  \bibinfo {author} {\bibfnamefont{W.}~\bibnamefont{Doster}},\ }%
  \bibfield{journal}{%
  \bibinfo {journal} {Biochim.\ Biophys.\ Acta}\ }%
  \textbf{\bibinfo {volume} {1804}},\ \bibinfo {pages} {3} (\bibinfo {year}
  {2010})%
  \bibAnnoteFile{NoStop}{DosterBBA:10}%
\bibitem{Fenimore:04}%
  \BibitemOpen
  \bibfield{author}{%
  \bibinfo {author} {\bibfnamefont{P.~W.}\ \bibnamefont{Fenimore}}, \bibinfo
  {author} {\bibfnamefont{H.}~\bibnamefont{Frauenfelder}}, \bibinfo {author}
  {\bibfnamefont{B.~H.}\ \bibnamefont{McMahon}},\ and\ \bibinfo {author}
  {\bibfnamefont{R.~D.}\ \bibnamefont{Young}},\ }%
  \bibfield{journal}{%
  \bibinfo {journal} {Proc. Natl. Acad. Sci.}\ }%
  \textbf{\bibinfo {volume} {101}},\ \bibinfo {pages} {14408} (\bibinfo {year}
  {2004})%
  \bibAnnoteFile{NoStop}{Fenimore:04}%
\bibitem{ChenPM:08}%
  \BibitemOpen
  \bibfield{author}{%
  \bibinfo {author} {\bibfnamefont{G.}~\bibnamefont{Chen}}, \bibinfo {author}
  {\bibfnamefont{P.~W.}\ \bibnamefont{Fenimore}}, \bibinfo {author}
  {\bibfnamefont{H.}~\bibnamefont{Frauenfelder}}, \bibinfo {author}
  {\bibfnamefont{F.}~\bibnamefont{Mezei}}, \bibinfo {author}
  {\bibfnamefont{J.}~\bibnamefont{Swenson}},\ and\ \bibinfo {author}
  {\bibfnamefont{R.~D.}\ \bibnamefont{Young}},\ }%
  \bibfield{journal}{%
  \bibinfo {journal} {Phil.\ Mag.}\ }%
  \textbf{\bibinfo {volume} {88}},\ \bibinfo {pages} {33} (\bibinfo {year}
  {2008})%
  \bibAnnoteFile{NoStop}{ChenPM:08}%
\bibitem{Frauenfelder:09}%
  \BibitemOpen
  \bibfield{author}{%
  \bibinfo {author} {\bibfnamefont{H.}~\bibnamefont{Frauenfelder}}, \bibinfo
  {author} {\bibfnamefont{G.}~\bibnamefont{Chen}}, \bibinfo {author}
  {\bibfnamefont{J.}~\bibnamefont{Berendzen}}, \bibinfo {author}
  {\bibfnamefont{P.~W.}\ \bibnamefont{Fenimore}}, \bibinfo {author}
  {\bibfnamefont{H.}~\bibnamefont{Jansson}}, \bibinfo {author}
  {\bibfnamefont{B.~H.}\ \bibnamefont{McMahon}}, \bibinfo {author}
  {\bibfnamefont{I.~R.}\ \bibnamefont{Stroe}}, \bibinfo {author}
  {\bibfnamefont{J.}~\bibnamefont{Swenson}},\ and\ \bibinfo {author}
  {\bibfnamefont{R.~D.}\ \bibnamefont{Young}},\ }%
  \bibfield{journal}{%
  \bibinfo {journal} {Proc.\ Natl.\ Acad.\ Sci.}\ }%
  \textbf{\bibinfo {volume} {106}},\ \bibinfo {pages} {5129} (\bibinfo {year}
  {2009})%
  \bibAnnoteFile{NoStop}{Frauenfelder:09}%
\bibitem{Ediger:96}%
  \BibitemOpen
  \bibfield{author}{%
  \bibinfo {author} {\bibfnamefont{M.~D.}\ \bibnamefont{Ediger}}, \bibinfo
  {author} {\bibfnamefont{C.~A.}\ \bibnamefont{Angell}},\ and\ \bibinfo
  {author} {\bibfnamefont{S.~R.}\ \bibnamefont{Nagel}},\ }%
  \bibfield{journal}{%
  \bibinfo {journal} {J.\ Phys.\ Chem.}\ }%
  \textbf{\bibinfo {volume} {100}},\ \bibinfo {pages} {13200} (\bibinfo {year}
  {1996})%
  \bibAnnoteFile{NoStop}{Ediger:96}%
\bibitem{HePRL:08}%
  \BibitemOpen
  \bibfield{author}{%
  \bibinfo {author} {\bibfnamefont{Y.}~\bibnamefont{He}}, \bibinfo {author}
  {\bibfnamefont{P.~I.}\ \bibnamefont{Ku}}, \bibinfo {author}
  {\bibfnamefont{J.~R.}\ \bibnamefont{Knab}}, \bibinfo {author}
  {\bibfnamefont{J.~Y.}\ \bibnamefont{Chen}},\ and\ \bibinfo {author}
  {\bibfnamefont{A.~G.}\ \bibnamefont{Markelz}},\ }%
  \bibfield{journal}{%
  \bibinfo {journal} {Phys.\ Rev.\ Lett.}\ }%
  \textbf{\bibinfo {volume} {101}},\ \bibinfo {eid} {178103} (\bibinfo {year}
  {2008})%
  \bibAnnoteFile{NoStop}{HePRL:08}%
\bibitem{Khodadadi:08}%
  \BibitemOpen
  \bibfield{author}{%
  \bibinfo {author} {\bibfnamefont{S.}~\bibnamefont{Khodadadi}}, \bibinfo
  {author} {\bibfnamefont{S.}~\bibnamefont{Pawlus}}, \bibinfo {author}
  {\bibfnamefont{J.~H.}\ \bibnamefont{Roh}}, \bibinfo {author}
  {\bibfnamefont{V.~G.}\ \bibnamefont{Sakai}}, \bibinfo {author}
  {\bibfnamefont{E.}~\bibnamefont{Mamontov}},\ and\ \bibinfo {author}
  {\bibfnamefont{A.~P.}\ \bibnamefont{Sokolov}},\ }%
  \bibfield{journal}{%
  \bibinfo {journal} {J.\ Chem.\ Phys.}\ }%
  \textbf{\bibinfo {volume} {128}},\ \bibinfo {pages} {195106} (\bibinfo {year}
  {2008})%
  \bibAnnoteFile{NoStop}{Khodadadi:08}%
\bibitem{Tarek_PhysRevLett:02}%
  \BibitemOpen
  \bibfield{author}{%
  \bibinfo {author} {\bibfnamefont{M.}~\bibnamefont{Tarek}}\ and\ \bibinfo
  {author} {\bibfnamefont{D.~J.}\ \bibnamefont{Tobias}},\ }%
  \bibfield{journal}{%
  \bibinfo {journal} {Phys. Rev. Lett.}\ }%
  \textbf{\bibinfo {volume} {88}},\ \bibinfo {pages} {138101} (\bibinfo {year}
  {2002})%
  \bibAnnoteFile{NoStop}{Tarek_PhysRevLett:02}%
\bibitem{Singh:81}%
  \BibitemOpen
  \bibfield{author}{%
  \bibinfo {author} {\bibfnamefont{G.~P.}\ \bibnamefont{Singh}}, \bibinfo
  {author} {\bibfnamefont{F.}~\bibnamefont{Parak}}, \bibinfo {author}
  {\bibfnamefont{S.}~\bibnamefont{Hunklinger}},\ and\ \bibinfo {author}
  {\bibfnamefont{K.}~\bibnamefont{Dransfeld}},\ }%
  \bibfield{journal}{%
  \bibinfo {journal} {Phys. Rev. Lett.}\ }%
  \textbf{\bibinfo {volume} {47}},\ \bibinfo {pages} {685} (\bibinfo {year}
  {1981})%
  \bibAnnoteFile{NoStop}{Singh:81}%
\bibitem{Magazu:10}%
  \BibitemOpen
  \bibfield{author}{%
  \bibinfo {author} {\bibfnamefont{S.}~\bibnamefont{Magaz{\`u}}}, \bibinfo
  {author} {\bibfnamefont{F.}~\bibnamefont{Migliardo}},\ and\ \bibinfo {author}
  {\bibfnamefont{A.}~\bibnamefont{Benedetto}},\ }%
  \bibfield{journal}{%
  \bibinfo {journal} {J.\ Phys.\ Chem. B}\ }%
  \textbf{\bibinfo {volume} {114}},\ \bibinfo {pages} {9268} (\bibinfo {year}
  {2010})%
  \bibAnnoteFile{NoStop}{Magazu:10}%
\bibitem{DMjcp1:09}%
  \BibitemOpen
  \bibfield{author}{%
  \bibinfo {author} {\bibfnamefont{D.~V.}\ \bibnamefont{Matyushov}},\ }%
  \bibfield{journal}{%
  \bibinfo {journal} {J.\ Chem.\ Phys.}\ }%
  \textbf{\bibinfo {volume} {130}},\ \bibinfo {pages} {164522} (\bibinfo {year}
  {2009})%
  \bibAnnoteFile{NoStop}{DMjcp1:09}%
\bibitem{Achterhold:02}%
  \BibitemOpen
  \bibfield{author}{%
  \bibinfo {author} {\bibfnamefont{K.}~\bibnamefont{Achterhold}}, \bibinfo
  {author} {\bibfnamefont{C.}~\bibnamefont{Keppler}}, \bibinfo {author}
  {\bibfnamefont{A.}~\bibnamefont{Ostermann}}, \bibinfo {author}
  {\bibfnamefont{U.}~\bibnamefont{van B{\"u}rck}}, \bibinfo {author}
  {\bibfnamefont{W.}~\bibnamefont{Sturhahn}}, \bibinfo {author}
  {\bibfnamefont{E.~E.}\ \bibnamefont{Alp}},\ and\ \bibinfo {author}
  {\bibfnamefont{F.~G.}\ \bibnamefont{Parak}},\ }%
  \bibfield{journal}{%
  \bibinfo {journal} {Phys. Rev. E}\ }%
  \textbf{\bibinfo {volume} {65}},\ \bibinfo {pages} {051916} (\bibinfo {year}
  {2002})%
  \bibAnnoteFile{NoStop}{Achterhold:02}%
\bibitem{Leu:08}%
  \BibitemOpen
  \bibfield{author}{%
  \bibinfo {author} {\bibfnamefont{B.~M.}\ \bibnamefont{Leu}}, \bibinfo
  {author} {\bibfnamefont{Y.}~\bibnamefont{Zhang}}, \bibinfo {author}
  {\bibfnamefont{L.}~\bibnamefont{Bu}}, \bibinfo {author}
  {\bibfnamefont{J.~E.}\ \bibnamefont{Straub}}, \bibinfo {author}
  {\bibfnamefont{J.}~\bibnamefont{Zhao}}, \bibinfo {author}
  {\bibfnamefont{W.}~\bibnamefont{Sturhahn}}, \bibinfo {author}
  {\bibfnamefont{E.~E.}\ \bibnamefont{Alp}},\ and\ \bibinfo {author}
  {\bibfnamefont{J.~T.}\ \bibnamefont{Sage}},\ }%
  \bibfield{journal}{%
  \bibinfo {journal} {Biophys. J.}\ }%
  \textbf{\bibinfo {volume} {95}},\ \bibinfo {pages} {5874} (\bibinfo {year}
  {2008})%
  \bibAnnoteFile{NoStop}{Leu:08}%
\bibitem{Leu:10}%
  \BibitemOpen
  \bibfield{author}{%
  \bibinfo {author} {\bibfnamefont{B.~M.}\ \bibnamefont{Leu}}, \bibinfo
  {author} {\bibfnamefont{A.}~\bibnamefont{Alatas}}, \bibinfo {author}
  {\bibfnamefont{H.}~\bibnamefont{Sinn}}, \bibinfo {author}
  {\bibfnamefont{E.~E.}\ \bibnamefont{Alp}}, \bibinfo {author}
  {\bibfnamefont{A.~H.}\ \bibnamefont{Said}}, \bibinfo {author}
  {\bibfnamefont{H.}~\bibnamefont{Yavas}}, \bibinfo {author}
  {\bibfnamefont{J.}~\bibnamefont{Zhao}}, \bibinfo {author}
  {\bibfnamefont{J.~T.}\ \bibnamefont{Sage}},\ and\ \bibinfo {author}
  {\bibfnamefont{W.}~\bibnamefont{Sturhahn}},\ }%
  \bibfield{journal}{%
  \bibinfo {journal} {J.\ Chem.\ Phys.}\ }%
  \textbf{\bibinfo {volume} {132}},\ \bibinfo {pages} {085103} (\bibinfo {year}
  {2010})%
  \bibAnnoteFile{NoStop}{Leu:10}%
\bibitem{Landau7}%
  \BibitemOpen
  \bibfield{author}{%
  \bibinfo {author} {\bibfnamefont{L.~D.}\ \bibnamefont{Landau}}\ and\ \bibinfo
  {author} {\bibfnamefont{E.~M.}\ \bibnamefont{Lifshits}},\ }%
  \emph{\bibinfo {title} {Theory of elasticity}}\ (\bibinfo {publisher}
  {Elsevier},\ \bibinfo {address} {Amsterdam},\ \bibinfo {year} {1986})\
  \bibinfo {note} {p. 18}%
  \bibAnnoteFile{NoStop}{Landau7}%
\bibitem{Christensen:03}%
  \BibitemOpen
  \bibfield{author}{%
  \bibinfo {author} {\bibfnamefont{R.~M.}\ \bibnamefont{Christensen}},\ }%
  \emph{\bibinfo {title} {Theory of Viscoelasticity}}\ (\bibinfo {publisher}
  {Dover Publications, Inc.},\ \bibinfo {address} {Mineola, N.\ Y.},\ \bibinfo
  {year} {2003})%
  \bibAnnoteFile{NoStop}{Christensen:03}%
\bibitem{Boon:91}%
  \BibitemOpen
  \bibfield{author}{%
  \bibinfo {author} {\bibfnamefont{J.~P.}\ \bibnamefont{Boon}}\ and\ \bibinfo
  {author} {\bibfnamefont{S.}~\bibnamefont{Yip}},\ }%
  \emph{\bibinfo {title} {Molecular Hydrodynamics}}\ (\bibinfo {publisher}
  {Dover Publications, Inc.},\ \bibinfo {address} {New York},\ \bibinfo {year}
  {1991})%
  \bibAnnoteFile{NoStop}{Boon:91}%
\bibitem{Singwi:60}%
  \BibitemOpen
  \bibfield{author}{%
  \bibinfo {author} {\bibfnamefont{K.~S.}\ \bibnamefont{Singwi}}\ and\ \bibinfo
  {author} {\bibfnamefont{A.}~\bibnamefont{Sjolander}},\ }%
  \bibfield{journal}{%
  \bibinfo {journal} {Phys.\ Rev.}\ }%
  \textbf{\bibinfo {volume} {120}},\ \bibinfo {pages} {1093} (\bibinfo {year}
  {1960})%
  \bibAnnoteFile{NoStop}{Singwi:60}%
\bibitem{Cranshaw:85}%
  \BibitemOpen
  \bibfield{author}{%
  \bibinfo {author} {\bibfnamefont{T.~E.}\ \bibnamefont{Cranshaw}}, \bibinfo
  {author} {\bibfnamefont{B.~W.}\ \bibnamefont{Dale}}, \bibinfo {author}
  {\bibfnamefont{G.~O.}\ \bibnamefont{Longworth}},\ and\ \bibinfo {author}
  {\bibfnamefont{C.~E.}\ \bibnamefont{Johnson}},\ }%
  \emph{\bibinfo {title} {M{\"o}ssbauer spectroscopy and its applications}}\
  (\bibinfo {publisher} {Cambridge University Press},\ \bibinfo {address}
  {Cambridge},\ \bibinfo {year} {1985})%
  \bibAnnoteFile{NoStop}{Cranshaw:85}%
\bibitem{ChaikinLubensky}%
  \BibitemOpen
  \bibfield{author}{%
  \bibinfo {author} {\bibfnamefont{P.~M.}\ \bibnamefont{Chaikin}}\ and\
  \bibinfo {author} {\bibfnamefont{T.~C.}\ \bibnamefont{Lubensky}},\ }%
  \emph{\bibinfo {title} {Principles of condensed matter physics}}\ (\bibinfo
  {publisher} {Cambridge University Press},\ \bibinfo {address} {Cambridge},\
  \bibinfo {year} {1995})%
  \bibAnnoteFile{NoStop}{ChaikinLubensky}%
\bibitem{Knapp:83}%
  \BibitemOpen
  \bibfield{author}{%
  \bibinfo {author} {\bibfnamefont{E.~W.}\ \bibnamefont{Knapp}}, \bibinfo
  {author} {\bibfnamefont{S.~F.}\ \bibnamefont{Fischer}},\ and\ \bibinfo
  {author} {\bibfnamefont{F.}~\bibnamefont{Parak}},\ }%
  \bibfield{journal}{%
  \bibinfo {journal} {J.\ Chem.\ Phys.}\ }%
  \textbf{\bibinfo {volume} {78}},\ \bibinfo {pages} {4701} (\bibinfo {year}
  {1983})%
  \bibAnnoteFile{NoStop}{Knapp:83}%
\bibitem{Mori:06}%
  \BibitemOpen
  \bibfield{author}{%
  \bibinfo {author} {\bibfnamefont{K.}~\bibnamefont{Mori}}, \bibinfo {author}
  {\bibfnamefont{Y.}~\bibnamefont{Seki}}, \bibinfo {author}
  {\bibfnamefont{Y.}~\bibnamefont{Yamada}},\ and\ \bibinfo {author}
  {\bibfnamefont{H.~M.~K.}\ \bibnamefont{Soda}},\ }%
  \bibfield{journal}{%
  \bibinfo {journal} {J.\ Chem.\ Phys.}\ }%
  \textbf{\bibinfo {volume} {125}},\ \bibinfo {pages} {054903} (\bibinfo {year}
  {2006})%
  \bibAnnoteFile{NoStop}{Mori:06}%
\bibitem{Harris:04}%
  \BibitemOpen
  \bibfield{author}{%
  \bibinfo {author} {\bibfnamefont{K.~R.}\ \bibnamefont{Harris}}\ and\ \bibinfo
  {author} {\bibfnamefont{L.~A.}\ \bibnamefont{Woolf}},\ }%
  \bibfield{journal}{%
  \bibinfo {journal} {J. Chem. Eng. Data}\ }%
  \textbf{\bibinfo {volume} {49}},\ \bibinfo {pages} {1064} (\bibinfo {year}
  {2004})%
  \bibAnnoteFile{NoStop}{Harris:04}%
\bibitem{Slie:66}%
  \BibitemOpen
  \bibfield{author}{%
  \bibinfo {author} {\bibfnamefont{W.~M.}\ \bibnamefont{Slie}}, \bibinfo
  {author} {\bibfnamefont{J.}~\bibnamefont{A.~R.~Donfor}},\ and\ \bibinfo
  {author} {\bibfnamefont{T.~A.}\ \bibnamefont{Litovitz}},\ }%
  \bibfield{journal}{%
  \bibinfo {journal} {J.\ Chem.\ Phys.}\ }%
  \textbf{\bibinfo {volume} {44}},\ \bibinfo {pages} {3712} (\bibinfo {year}
  {1966})%
  \bibAnnoteFile{NoStop}{Slie:66}%
\bibitem{Monaco:99}%
  \BibitemOpen
  \bibfield{author}{%
  \bibinfo {author} {\bibfnamefont{G.}~\bibnamefont{Monaco}}, \bibinfo {author}
  {\bibfnamefont{A.}~\bibnamefont{Cunsolo}}, \bibinfo {author}
  {\bibfnamefont{G.}~\bibnamefont{Ruocco}},\ and\ \bibinfo {author}
  {\bibfnamefont{F.}~\bibnamefont{Sette}},\ }%
  \bibfield{journal}{%
  \bibinfo {journal} {Phys. Rev. E}\ }%
  \textbf{\bibinfo {volume} {60}},\ \bibinfo {pages} {5505} (\bibinfo {year}
  {1999})%
  \bibAnnoteFile{NoStop}{Monaco:99}%
\bibitem{Kharakoz:00}%
  \BibitemOpen
  \bibfield{author}{%
  \bibinfo {author} {\bibfnamefont{D.~P.}\ \bibnamefont{Kharakoz}},\ }%
  \bibfield{journal}{%
  \bibinfo {journal} {Biophys. J.}\ }%
  \textbf{\bibinfo {volume} {79}},\ \bibinfo {pages} {511} (\bibinfo {year}
  {2000})%
  \bibAnnoteFile{NoStop}{Kharakoz:00}%
\bibitem{Morozov:85}%
  \BibitemOpen
  \bibfield{author}{%
  \bibinfo {author} {\bibfnamefont{V.~N.}\ \bibnamefont{Morozov}}\ and\
  \bibinfo {author} {\bibfnamefont{S.~G.}\ \bibnamefont{Gevorkian}},\ }%
  \bibfield{journal}{%
  \bibinfo {journal} {Biopolymers}\ }%
  \textbf{\bibinfo {volume} {24}},\ \bibinfo {pages} {1785} (\bibinfo {year}
  {1985})%
  \bibAnnoteFile{NoStop}{Morozov:85}%
\bibitem{Morozov:93}%
  \BibitemOpen
  \bibfield{author}{%
  \bibinfo {author} {\bibfnamefont{V.~N.}\ \bibnamefont{Morozov}}\ and\
  \bibinfo {author} {\bibfnamefont{Y.~Y.}\ \bibnamefont{Morozova}},\ }%
  \bibfield{journal}{%
  \bibinfo {journal} {J.\ Biomol.\ Struct.\ Dyn.}\ }%
  \textbf{\bibinfo {volume} {11}},\ \bibinfo {pages} {459} (\bibinfo {year}
  {1993})%
  \bibAnnoteFile{NoStop}{Morozov:93}%
\bibitem{DMunpubl:11}%
  \BibitemOpen
  \bibfield{author}{%
  \bibinfo {author} {\bibfnamefont{D.~V.}\ \bibnamefont{Matyushov}},\ }%
  \bibinfo {note} {unpublished}%
  \bibAnnoteFile{NoStop}{DMunpubl:11}%
\bibitem{DMjcp1:04}%
  \BibitemOpen
  \bibfield{author}{%
  \bibinfo {author} {\bibfnamefont{D.~V.}\ \bibnamefont{Matyushov}},\ }%
  \bibfield{journal}{%
  \bibinfo {journal} {J.\ Chem.\ Phys.}\ }%
  \textbf{\bibinfo {volume} {120}},\ \bibinfo {pages} {1375} (\bibinfo {year}
  {2004})%
  \bibAnnoteFile{NoStop}{DMjcp1:04}%
\bibitem{Swenson:06}%
  \BibitemOpen
  \bibfield{author}{%
  \bibinfo {author} {\bibfnamefont{J.}~\bibnamefont{Swenson}}, \bibinfo
  {author} {\bibfnamefont{H.}~\bibnamefont{Jansson}},\ and\ \bibinfo {author}
  {\bibfnamefont{R.}~\bibnamefont{Bergman}},\ }%
  \bibfield{journal}{%
  \bibinfo {journal} {Phys. Rev. Lett.}\ }%
  \textbf{\bibinfo {volume} {96}},\ \bibinfo {pages} {247802} (\bibinfo {year}
  {2006})%
  \bibAnnoteFile{NoStop}{Swenson:06}%
\bibitem{Schiro:09}%
  \BibitemOpen
  \bibfield{author}{%
  \bibinfo {author} {\bibfnamefont{G.}~\bibnamefont{Schir{\`o}}}, \bibinfo
  {author} {\bibfnamefont{A.}~\bibnamefont{Cupane}}, \bibinfo {author}
  {\bibfnamefont{E.}~\bibnamefont{Vitrano}},\ and\ \bibinfo {author}
  {\bibfnamefont{F.}~\bibnamefont{Bruni}},\ }%
  \bibfield{journal}{%
  \bibinfo {journal} {J.\ Phys.\ Chem. B}\ }%
  \textbf{\bibinfo {volume} {113}},\ \bibinfo {pages} {9606} (\bibinfo {year}
  {2009})%
  \bibAnnoteFile{NoStop}{Schiro:09}%
\bibitem{Bonura:10}%
  \BibitemOpen
  \bibfield{author}{%
  \bibinfo {author} {\bibfnamefont{M.}~\bibnamefont{Bonura}}, \bibinfo {author}
  {\bibfnamefont{G.}~\bibnamefont{Schir{\`o}}},\ and\ \bibinfo {author}
  {\bibfnamefont{A.}~\bibnamefont{Cupane}},\ }%
  \bibfield{journal}{%
  \bibinfo {journal} {Spectroscopy}\ }%
  \textbf{\bibinfo {volume} {24}},\ \bibinfo {pages} {143} (\bibinfo {year}
  {2010})%
  \bibAnnoteFile{NoStop}{Bonura:10}%
\bibitem{DMjpcb2:08}%
  \BibitemOpen
  \bibfield{author}{%
  \bibinfo {author} {\bibfnamefont{D.~N.}\ \bibnamefont{LeBard}}, \bibinfo
  {author} {\bibfnamefont{V.}~\bibnamefont{Kapko}},\ and\ \bibinfo {author}
  {\bibfnamefont{D.~V.}\ \bibnamefont{Matyushov}},\ }%
  \bibfield{journal}{%
  \bibinfo {journal} {J. Phys. Chem. B}\ }%
  \textbf{\bibinfo {volume} {112}},\ \bibinfo {pages} {10322} (\bibinfo {year}
  {2008})%
  \bibAnnoteFile{NoStop}{DMjpcb2:08}%
\bibitem{Lusceac:10}%
  \BibitemOpen
  \bibfield{author}{%
  \bibinfo {author} {\bibfnamefont{S.~A.}\ \bibnamefont{Lusceac}}, \bibinfo
  {author} {\bibfnamefont{M.~R.}\ \bibnamefont{Vogel}},\ and\ \bibinfo {author}
  {\bibfnamefont{C.~R.}\ \bibnamefont{Herbers}},\ }%
  \bibfield{journal}{%
  \bibinfo {journal} {Biochim.\ Biophys.\ Acta}\ }%
  \textbf{\bibinfo {volume} {1804}},\ \bibinfo {pages} {41} (\bibinfo {year}
  {2010})%
  \bibAnnoteFile{NoStop}{Lusceac:10}%
\bibitem{Doster:10}%
  \BibitemOpen
  \bibfield{author}{%
  \bibinfo {author} {\bibfnamefont{W.}~\bibnamefont{Doster}}, \bibinfo {author}
  {\bibfnamefont{S.}~\bibnamefont{Busch}}, \bibinfo {author}
  {\bibfnamefont{A.~M.}\ \bibnamefont{Gaspar}}, \bibinfo {author}
  {\bibfnamefont{M.-S.}\ \bibnamefont{Appavou}}, \bibinfo {author}
  {\bibfnamefont{J.}~\bibnamefont{Wuttke}},\ and\ \bibinfo {author}
  {\bibfnamefont{H.}~\bibnamefont{Scheer}},\ }%
  \bibfield{journal}{%
  \bibinfo {journal} {Phys. Rev. Lett.}\ }%
  \textbf{\bibinfo {volume} {104}},\ \bibinfo {pages} {098101} (\bibinfo {year}
  {2010})%
  \bibAnnoteFile{NoStop}{Doster:10}%
\bibitem{Khodadadi:10}%
  \BibitemOpen
  \bibfield{author}{%
  \bibinfo {author} {\bibfnamefont{S.}~\bibnamefont{Khodadadi}}, \bibinfo
  {author} {\bibfnamefont{A.}~\bibnamefont{Malkovskiy}}, \bibinfo {author}
  {\bibfnamefont{A.}~\bibnamefont{Kisliuk}},\ and\ \bibinfo {author}
  {\bibfnamefont{A.~P.}\ \bibnamefont{Sokolov}},\ }%
  \bibfield{journal}{%
  \bibinfo {journal} {Biochim.\ Biophys.\ Acta}\ }%
  \textbf{\bibinfo {volume} {1804}},\ \bibinfo {pages} {15} (\bibinfo {year}
  {2010})%
  \bibAnnoteFile{NoStop}{Khodadadi:10}%
\bibitem{Daniel:02}%
  \BibitemOpen
  \bibfield{author}{%
  \bibinfo {author} {\bibfnamefont{R.~M.}\ \bibnamefont{Daniel}}, \bibinfo
  {author} {\bibfnamefont{J.~L.}\ \bibnamefont{Finney}},\ and\ \bibinfo
  {author} {\bibfnamefont{J.~C.}\ \bibnamefont{Smith}},\ }%
  \bibfield{journal}{%
  \bibinfo {journal} {Faraday Discuss.}\ }%
  \textbf{\bibinfo {volume} {122}},\ \bibinfo {pages} {163} (\bibinfo {year}
  {2002})%
  \bibAnnoteFile{NoStop}{Daniel:02}%
\bibitem{Krishnan:08}%
  \BibitemOpen
  \bibfield{author}{%
  \bibinfo {author} {\bibfnamefont{M.}~\bibnamefont{Krishnan}}, \bibinfo
  {author} {\bibfnamefont{V.}~\bibnamefont{Kurkal-Siebert}},\ and\ \bibinfo
  {author} {\bibfnamefont{J.}~\bibnamefont{Smith}},\ }%
  \bibfield{journal}{%
  \bibinfo {journal} {J.\ Phys.\ Chem. B}\ }%
  \textbf{\bibinfo {volume} {112}},\ \bibinfo {pages} {5522} (\bibinfo {year}
  {2008})%
  \bibAnnoteFile{NoStop}{Krishnan:08}%
\bibitem{DMpre2:08}%
  \BibitemOpen
  \bibfield{author}{%
  \bibinfo {author} {\bibfnamefont{D.~N.}\ \bibnamefont{LeBard}}\ and\ \bibinfo
  {author} {\bibfnamefont{D.~V.}\ \bibnamefont{Matyushov}},\ }%
  \bibfield{journal}{%
  \bibinfo {journal} {Phys. Rev. E}\ }%
  \textbf{\bibinfo {volume} {78}},\ \bibinfo {pages} {061901} (\bibinfo {year}
  {2008})%
  \bibAnnoteFile{NoStop}{DMpre2:08}%
\bibitem{DMjpcb:10}%
  \BibitemOpen
  \bibfield{author}{%
  \bibinfo {author} {\bibfnamefont{D.~N.}\ \bibnamefont{LeBard}}\ and\ \bibinfo
  {author} {\bibfnamefont{D.~V.}\ \bibnamefont{Matyushov}},\ }%
  \bibfield{journal}{%
  \bibinfo {journal} {J.\ Phys.\ Chem. B}\ }%
  \textbf{\bibinfo {volume} {114}},\ \bibinfo {pages} {9246} (\bibinfo {year}
  {2010})%
  \bibAnnoteFile{NoStop}{DMjpcb:10}%
\bibitem{Frolov:97}%
  \BibitemOpen
  \bibfield{author}{%
  \bibinfo {author} {\bibfnamefont{E.~N.}\ \bibnamefont{Frolov}}, \bibinfo
  {author} {\bibfnamefont{R.}~\bibnamefont{Gvosdev}}, \bibinfo {author}
  {\bibfnamefont{V.~I.}\ \bibnamefont{Goldanskii}},\ and\ \bibinfo {author}
  {\bibfnamefont{F.~G.}\ \bibnamefont{Parak}},\ }%
  \bibfield{journal}{%
  \bibinfo {journal} {J. Biol. Inorg. Chem.}\ }%
  \textbf{\bibinfo {volume} {2}},\ \bibinfo {pages} {710} (\bibinfo {year}
  {1997})%
  \bibAnnoteFile{NoStop}{Frolov:97}%
\bibitem{Jorgenson:05}%
  \BibitemOpen
  \bibfield{author}{%
  \bibinfo {author} {\bibfnamefont{A.~M.}\ \bibnamefont{Jorgens{\o}n}},
  \bibinfo {author} {\bibfnamefont{F.}~\bibnamefont{Parak}},\ and\ \bibinfo
  {author} {\bibfnamefont{H.~E.~M.}\ \bibnamefont{Christensen}},\ }%
  \bibfield{journal}{%
  \bibinfo {journal} {Phys.\ Chem.\ Chem.\ Phys.}\ }%
  \textbf{\bibinfo {volume} {7}},\ \bibinfo {pages} {3472} (\bibinfo {year}
  {2005})%
  \bibAnnoteFile{NoStop}{Jorgenson:05}%
\end{thebibliography}%

\end{document}